\documentclass[floats,amsmath,amssymb,a4paper,noshowpacs,twocolumn,prb]{revtex4}%
\usepackage{graphicx}
\usepackage{amsmath}
\usepackage{amsfonts}
\usepackage{amssymb}
\usepackage{dcolumn}%
\providecommand{\U}[1]{\protect\rule{.1in}{.1in}}

\begin{document}
\title{Thermoelectric Effects in Magnetic Nanostructures}
\author{Moosa Hatami and Gerrit E. W. Bauer}
\affiliation{Kavli Institute of NanoScience, Delft University of Technology, Lorentzweg 1,
2628 CJ Delft, The Netherlands}
\author{Qinfang Zhang and Paul J. Kelly}
\affiliation{Faculty of Science and Technology and MESA$^{+}$ Institute for Nanotechnology,
University of Twente, P.O. Box 217, 7500 AE Enschede, The Netherlands}
\date{\today}

\begin{abstract}
We model and evaluate the Peltier and Seebeck effects in magnetic multilayer
nanostructures by a finite-element theory of thermoelectric properties. We
present analytical expressions for the thermopower and the current-induced
temperature changes due to Peltier cooling/heating. The thermopower
of a magnetic element is in general spin-polarized, leading to spin-heat coupling effects. 
Thermoelectric effects in spin valves depend on the relative alignment 
of the magnetization directions and are sensitive to spin-flip scattering as well as 
inelastic collisions in the normal metal spacer.

\end{abstract}
\maketitle

\section{Introduction}

The Peltier effect refers to the conversion of an electric voltage into a
temperature difference (that can be used for refrigeration), while the Seebeck
effect refers to the inverse process, the generation of an electric field by a
temperature gradient.\cite{Giazotto:rmp06} Renewed interest in thermoelectric
properties is motivated in part by the improved performance of nanometer-scale
structures.\cite{Hicks:prb93, Sales:sc02,Boukai:nat08,Hochbaum:nat08}
Thin-film thermoelectric coolers can provide cheap and fast spot-cooling in
micro- and nanoelectronic circuits and
devices.\cite{Venkatasubramanian:nat01,Ohta:natm07} Strongly enhanced
thermopower in quantum point contacts with widths approaching the Fermi
wavelength can be used for sensitive and local electron thermometry.
\cite{Molenkamp:prl92,vanHouten:sst92} In ferromagnets and heterostructures
involving magnetic elements, the effect of the magnetization (spin) degree of
freedom on thermoelectric transport has to be taken into
account.\cite{Johnson:prb87,Johnson:js03,Wegrowe:prb00} The giant
magneto-thermoelectric power in multilayered
nanopillars,\cite{Gravier:prb06a,Gravier:prb06b} thermally excited
spin-currents in metals with embedded ferromagnetic clusters
\cite{Serrano:natm06,Tsyplyatyev:prb06} and thermal spin-transfer torque in
spin-valve devices\cite{Hatami:prl07} are examples of spin-dependent
thermoelectric phenomena on a nanometer scale.

Recently a large Peltier effect was discovered in transition metal
multilayered nanopillars by Fukushima \textit{et al}%
.\cite{Fukushima:jjap05,Fukushima:ieeem05} The temperature and energy
dissipation as a function of an applied current were monitored using the
temperature-dependent electrical resistance. In asymmetric structures the
parabolic dependence of the resistance arising from current-induced Joule
heating was found to be modified by a superimposed linear (Peltier) term that
shifts the minimum resistance to a finite value of the current that could be
positive or negative, depending on the combination of materials. Gravier
\textit{et al.} \cite{Gravier:jpd06} used a model of diffuse thermoelectric
transport in (non-magnetic) metallic heterostructures to compute the Peltier
effect. These calculations had to be carried out numerically and magnetism was
not taken into account. The sample cross-sections were used as fitting
parameters that appeared to be too small compared to the actual sample
sizes.\cite{Fukushima:jjap05,Fukushima:ieeem05} This discrepancy was
attributed to the neglect of interface scattering. Katayama-Yoshida \textit{et
al}.\cite{Katayama:jjap07} interpreted the perceived enhancement of the
cooling power as a contribution from an adiabatic spin-entropy expansion term
$(k_{B}/e)\ln2$. Dubi and Di Ventra\cite{Dubi:09} studied the Seebeck effect in single level
quantum dots with ferromagnetic contacts. 

Enhancing the performance of solid state cooling elements remains a challenge
both for theory and experiment.
\cite{Giazotto:rmp06,Hicks:prb93,Sales:sc02,Boukai:nat08,Hochbaum:nat08,Venkatasubramanian:nat01,Ohta:natm07}
Fukushima \textit{et al}. \cite{Fukushima:jjap05,Fukushima:ieeem05} suggested
that the Peltier effect in transition metal nanostructures could be useful for
cooling magnetoelectronic devices. In order to assess this idea, the material
dependence of the Peltier effect in magnetic nanostructures has to be
understood. In this paper we investigate the Peltier effect in magnetic
heterostructures theoretically, taking into account spin-dependent interface
and bulk scattering by means of an extended finite element (circuit) theory of
transport\cite{Hatami:prl07} that is a generalization of magnetoelectronic
circuit theory.\cite{Brataas:prl00,Brataas:epjb01,Brataas:prp06} Such a theory
is also suitable to study the magneto-thermoelectric power in magnetic
multilayers in which transport is normal to the
interfaces.\cite{Gravier:prb06a,Gravier:prb06b}
\begin{figure}[t]
\resizebox{\columnwidth}{!} {\includegraphics{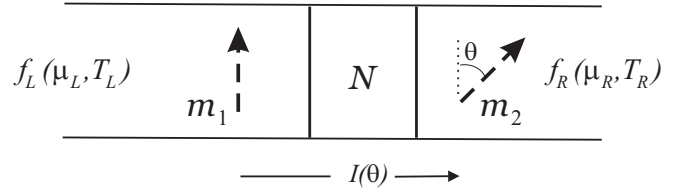}} \caption{
Schematic of a spin-valve structure connected to reservoirs at different
chemical potentials and/or temperatures ($f_{L(R)}$ are the Fermi-Dirac distribution functions). 
Charge and heat flows through the normal metal spacer (N) are functions of 
the angle between the magnetization directions ($\theta$) in the ferromagnets. 
}\label{fig1}%
\end{figure}

In the following we explain the method, initially disregarding bulk impurity
scattering, which is justified in the thin layer limit when interface
scattering is dominant. We start with a simple normal metal $N_{1}|N|N_{2}$
trilayer structure in which metal $N$ is sandwiched between two electron
reservoirs consisting of different metals $N_{1(2)}$. We then extend the
methodology by making first one and then both reservoirs magnetic. In the
$F_{1}|N|F_{2}$ (ferromagnet$|$normal-metal$|$ferromagnet) spin-valve sketched
in Fig.~\ref{fig1}, the relative orientation of the magnetization directions
as well as the strength of the inelastic collisions become important
parameters. Finally, we show how bulk impurity scattering (normal as well as spin-flip)
can be introduced into the formalism, discuss the relevance of our
results for experiments and finish with a summary and conclusions.

\section{Current-induced electron cooling and heating}

A resistor network theory is an efficient way to describe the electric
transport properties of magnetic heterostructures. 
Used to model the giant magnetoresistance effect in terms of the two-channel 
series resistor model, \cite{Valet:prb93,Bass:jmmm99} it has been generalized to 
include the spin transfer torque in non-collinear magnetization 
configurations.\cite{Brataas:prp06} A resistor
model for thermoelectric effects was introduced by
MacDonald\cite{MacDonald:62} to understand the effects of different types of
impurities in homogeneous bulk metals.

In the present section we develop a generalized (spin-less) circuit theory to
describe the effects of temperature and voltage bias on normal metal
heterostructures. Our starting point is a (non-unique) definition of the
circuit or device topology by partitioning it into reservoirs, resistors and
nodes. Discrete resistive elements are interfaces, potential barriers or
constrictions that limit the transport. We require that these resistors can be
represented by purely elastic scattering processes. The resistors are
separated by nodes, in which electrons can be described by semiclassical
distribution functions, $f_{i}$ in node $i$. When inelastic electron-electron
or electron-phonon interactions in the nodes are strong enough, the $f_{i}$
become Fermi-Dirac distributions parametrized by temperatures $T_{i}$ and
chemical potentials $\mu_{i}$. The charge and heat currents through a given
resistor, denoted by $I=\int d\epsilon\imath(\epsilon)$ and $e\dot{Q}=\int
d\epsilon(\epsilon-\mu)\imath(\epsilon)=eI^{E}-\mu I,$ respectively, where
$\mu$ is the global chemical potential in equilibrium, are determined via the
spectral current density
\begin{equation}
\imath(\epsilon)=G(\epsilon)[f_{L}(\epsilon)-f_{R}(\epsilon)],
\label{spectralcurrent}
\end{equation}
where $G(\epsilon)$ is an energy-dependent (spectral) conductance between two
neighboring nodes with distribution functions $f_{L(R)}$. According to
Landauer-B$\ddot{\text{u}}$ttiker scattering theory the conductance
\begin{equation}
G(\epsilon)=\frac{2e^{2}}{h}\sum_{mn}[\delta_{mn}-r_{mn}^{\ast}(\epsilon
)r_{mn}(\epsilon)]. \label{G}%
\end{equation}
depends on the energy-dependent reflection amplitudes $r_{mn}(\epsilon)$ at
the resistor for electrons that are incident from a node or reservoir. 
$2e^{2}/h \simeq 1/13\operatorname{k\Omega}$ is the quantum of conductance.
For structure elements such as interfaces which are not overly complicated,
$r_{mn}(\epsilon)$ can be calculated from microscopic, first-principles
calculations.\cite{Xia:prb06,Hatami:prl07,Zhang:prb09} Thermoelectric effects
arise from the energy dependence of $G(\epsilon)$. The circuit theory approach
requires that scattering in the resistive elements is elastic. $G(\epsilon)$
may in principle be bias dependent, which becomes important for tunnel
junctions. Here we concentrate on abrupt intermetallic interfaces with
bias-independent spectral conductances. The interface resistance of
transparent interfaces in a diffuse environment is affected by a correction
caused by the drift of the distribution
function.\cite{Schep:prb97,Brataas:prp06} The bare spectral resistance
$G(\epsilon)^{-1}$ is then substituted by $G(\epsilon)^{-1}- \left[  G_{N_{L}}(\epsilon)^{-1}+G_{N_{R}}(\epsilon)^{-1}\right]  $,
where $G_{N_{i}}(\epsilon)=2e^{2}M_{N_{i}}(\epsilon)/h$ 
are the Sharvin conductances of the
metals that form the interface and $M_{N_{i}}$ are 
the total numbers of single-spin transport modes.

Let us now consider a non-equilibrium steady state in a simple $N_{1}|N|N_{2}$
normal metal structure in which the chemical potential $\mu_{1(2)}%
=\mu-eV_{1(2)}$ and temperature $T_{1(2)}$ of the nodes deviate from their
equilibrium values $(\mu,T)$. Node $N$ is assumed to be fully thermalized with
a distribution described by $(eV_{N},T_{N})$ that still have to be determined.
To lowest order in the applied thermoelectric fields, we use the expansions
$f_{i}\approx f_{0}+\frac{\partial f_{0}}{\partial\mu}(\mu_{i}-\mu
)+\frac{\partial f_{0}}{\partial T}(T_{i}-T)$ in the following, where $f_{0}$ is the
equilibrium Fermi-Dirac distribution function. It is possible to proceed
and compute thermoelectric properties for arbitrary energy dependences of
$G(\epsilon)$. However, results become much simpler upon using the Sommerfeld
expansion \cite{Ashcroft:76} in $k_{B}T/\epsilon_{F}$. Provided the
conductances do not vary too rapidly near the Fermi energy
(to be precise $\partial_{\epsilon}G|_{\epsilon_{F}}\neq\infty$ and
$G(\epsilon)e^{-\epsilon/k_{B}T}|_{\epsilon\gg\epsilon_{F}}\rightarrow0$), the
following approximation is applied
\begin{align}
& \int_{-\infty}^{\infty}d\xi\left(  -\frac{\partial f_{0}}{\partial\xi}\right)
\xi^{n}G(\xi)\approx G(0)\delta_{n,0}+\frac{\pi^{2}}{6}(k_{B}T)^{2}\nonumber\\
& \times \left[n(n-1)\xi^{n-2}G(\xi)+2n\xi^{n-1}\partial_{\xi}G(\xi)+ \xi^{n}\partial
^{2}_{\xi}G(\xi)\right]  _{\xi=0} \label{S-int}%
\end{align}
where $\xi$ is the energy variable relative to the Fermi energy. 
Applying the Sommerfeld expansion to the expressions 
for the charge and heat currents in terms of the spectral current density, 
Eq.(\ref{spectralcurrent}), results in expressions for the 
charge and heat currents into the normal node through junctions 1 and 2 
, valid for $\frac{\pi^{2}}{6}(k_{B}T)^{2}\partial^{2}_{\epsilon
}G|_{\epsilon_{F}}\ll G|_{\epsilon_{F}}$:
\begin{equation}
\left(
\begin{array}
[c]{c}%
I_{1(2)}\\
\dot{Q}_{1(2)}%
\end{array}
\right)  =G_{1(2)}\left(
\begin{array}
[c]{cc}%
1 & S_{1(2)}\\
-S_{1(2)}T & -\mathcal{L}_{0}T
\end{array}
\right)  \left(
\begin{array}
[c]{c}%
V_{N}-V_{1(2)}\\
T_{N}-T_{1(2)}%
\end{array}
\right)  , \label{IQ}%
\end{equation}
where $\mathcal{L}_{0}=(k_{B}/e)^{2}\pi^{2}/3\simeq2.45\times10^{-8}%
\mathrm{V}^2\mathrm{K}^{-2}$ is the Lorenz number. $G_{i}=G_{i}\left(
\epsilon_{F}\right)  $ are the conductances and $S_{i}=-e\mathcal{L}%
_{0}T\partial_{\epsilon}\ln G_{i}|_{\epsilon_{F}}$ (Mott's formula) are the Seebeck
coefficients or thermopowers at the zero-temperature chemical potential, which
for metals is just the Fermi energy $\epsilon_{F}$. In bulk materials the
thermopower can be positive or negative and even change sign as a function of
the temperature.\cite{Colquitt:prb71} 

The chemical potential shift $eV_{N}$ and temperature $T_{N}$ of the central
normal node and thus the thermal distribution function $f_{N}$ are determined
by conservation laws: the charge and energy flows of the electrons are
conserved, $I_{1}+I_{2}=0$ and $\dot{Q}_{1}+\dot{Q}_{2}=0$. The latter is
affected in principle by the phonon heat conduction through the contacts (see
Appendix A) but is disregarded here since thermal transport in good metals is
dominated by the conduction electrons.\cite{Gundrum:prb05} 
Using Eq. (\ref{IQ}) and the conservation laws we find  
to lowest order in the charge current $I$ that the electron temperature $T_{N}$ of the normal island
is modified from the zero-charge current value $T_0$ as
\begin{equation}
T_{N}=T_{0}+\frac{(\Pi_{1}-\Pi_{2})I}{\kappa_{1}+\kappa_{2}}, \label{TN1}%
\end{equation}
The electron cooling or heating of asymmetric structures $(\Pi_{1}\neq\Pi
_{2})$ by the applied current is the Peltier effect. $\Pi_{1(2)}=S_{1(2)}T$
are the Peltier coefficients and
\begin{equation}
\kappa_{1(2)}=\mathcal{L}_{0}TG_{1(2)}(1-S_{1(2)}^{2}/\mathcal{L}_{0})
\end{equation}
are the thermal conductances of the resistive elements. Large thermopowers violate
the simple proportionality between electrical and heat conductance,
$\kappa=\mathcal{L}_{0}TG$, the Wiedemann-Franz law. The expression for the
zero-current temperature in the central node
\begin{equation}
T_{0}=\frac{\kappa_{1}T_{1}+\kappa_{2}T_{2}}{\kappa_{1}+\kappa_{2}},
\label{T0}%
\end{equation}
follows from energy conservation. The charge current $I=I_{\Delta V}+I_{\Delta
T}$ is excited by a voltage difference $\Delta V=V_{2}-V_{1}$ as well as the temperature
bias $\Delta T=T_{2}-T_{1}$. The total electric charge and heat currents are
then given by,
\begin{align}
I  &  =\bar{G}(\Delta V+\bar{S}\Delta T)\label{electric}\\
\dot{Q}  &  =-\bar{\Pi}I-\bar{\kappa}\Delta T,
\end{align}
where the total conductance becomes
\begin{equation}
\bar{G}=\frac{G_{1}G_{2}(\kappa_{1}+\kappa_{2})}{G_{1}\kappa_{1}+(G_{1}%
+G_{2})\kappa_{12}+G_{2}\kappa_{2}}\label{Rsum}
\end{equation}
with
\begin{equation}
\kappa_{12}\equiv2\mathcal{L}_{0}T~\frac{G_{1}G_{2}}{G_{1}+G_{2}}\left(
1-\frac{S_{1}S_{2}}{\mathcal{L}_{0}}\right)  .
\end{equation}
The total conductance/resistance ($\bar{R}=1/\bar{G}$) violates the series
resistor rule, $\bar{R}\neq R_{1}+R_{2}$ $\left(  \text{with }R_{i}%
=1/G_{i}\right)  $, which is only recovered when either $S_{1}=S_{2}$ or when
$S_{i}^{2}\ll\mathcal{L}_{0}$; the latter is always the case at low
temperatures. On the other hand, the total thermopower $\bar{S}$ 
(or Peltier coefficient $\bar{\Pi}=\bar{S}T$) and thermal
conductance $\bar{\kappa}$ do obey simple sum rules
\begin{align}
\frac{\bar{S}}{{\bar{\kappa}}}  &  =\frac{S_{1}}{\kappa_{1}}+\frac{S_{2}%
}{\kappa_{2}},\label{ssum}\\
\frac{1}{{\bar{\kappa}}}  &  =\frac{1}{\kappa_{1}}+\frac{1}{\kappa_{2}}.
\label{ksum}
\end{align}
When in series, two thermal or electrical resistances are additive, as expressed in Eq. (\ref{ksum})
or $\bar{R}=R_{1}+R_{2}$ which is valid in the limit $S_{i}^{2}\ll \mathcal{L}_{0}$.  
In contrast, the thermopower, Eq. (\ref{ssum}), depends on the spatial distribution 
of the scattering objects rather than its integral (the
thermopower in a homogeneous bulk metal does not depend on its length). Equation
(\ref{ssum}) holds not only for the spatially distributed scatterers
considered here, but can also describe the relative contributions of different
types of scatterers to the thermopower in bulk materials.\cite{MacDonald:62}

The second law of thermodynamics (a non-negative entropy production) requires
$\kappa>0$, which in the Sommerfeld approximation leads to\cite{Guttman:prb95}
$|S_{max}|=\sqrt{\mathcal{L}_{0}}\simeq157%
\operatorname{\mu V}%
\mathrm{/}\mathrm{%
\operatorname{K}%
}$. Defining the thermoelectric figure of merit $ZT=|{\bar{S}}\Delta
T_{N}/\Delta V|$, Eq. (\ref{TN1}) for the temperature change $\Delta
T_{N}=T_{N}-T_{0}$ induced by an applied voltage results in
\begin{equation}
ZT=\frac{1}{4}\left|\frac{\bar{S}(S_{1}-S_{2})}{\mathcal{L}_{0}-\bar{S}^{2}}\right|%
\end{equation}
assuming $G_{1}=G_{2}$ and $\left(  S_{1}-S_{2}\right)  S_{1(2)}\ll
\mathcal{L}_{0},$ so that $|\bar{S}_{max}|=\sqrt{\mathcal{L}_{0}}$ corresponds
to the maximum efficiency $ZT=\infty$. Even for the best thin film
thermoelectric materials\cite{Sales:sc02} $ZT\lessapprox3$, and for most
metallic structures $S^{2}\ll\mathcal{L}_{0}$. Quantum point contacts,
however, have larger thermopowers due to 
size quantization,\cite{Molenkamp:prl92,vanHouten:sst92} so $S^{2}$ \emph{can} be
comparable to $\mathcal{L}_{0}$ and the predicted effects should be observable
in nanoscale structures.

Equation (\ref{TN1}) holds for low current densities, for which $\delta
V/2\ll\Pi\ll2\mathcal{L}_{0}T^{2}/\delta V,$ where $\delta V$ is the voltage
drop over a single contact. Non-linear heating by applied
currents can be included my means of the quadratic term in the expansions of 
the distribution functions, \textit{i.e.}  
$f_{i}\approx f_{0}+\frac{\partial f_{0}}{\partial\mu}(\mu_{i}-\mu
)+\frac{\partial f_{0}}{\partial T}(T_{i}-T)+\frac{1}{2}
\frac{\partial^{2} f_{0}}{\partial\mu^{2}}(\mu_{i}-\mu)^{2}+ ...$, 
and using the approximation
$\int_{-\infty}^{\infty}d\xi\left( \frac{\partial^2 f_{0}}{\partial\xi^2}\right)
\xi G(\xi)\approx G(0)$. 
When $\Pi\ll\delta V\ll \mathcal{L}_{0}T^{2}/\Pi$, 
expressions for the nonlinear currents reduce to
\begin{eqnarray}
&& \left(
\begin{array}
[c]{c}%
I_{1(2)}\\
\dot{Q}_{1(2)}%
\end{array}
\right)  \simeq G_{1(2)}\left(
\begin{array}
[c]{cc}%
1 & S_{1(2)}\\
-\Pi_{1(2)}-\tilde{V}_{1(2)} & -\mathcal{L}_{0}\tilde{T}_{1(2)}%
\end{array}
\right)  \nonumber\\
&& \times \left(
\begin{array}
[c]{c}%
V_{N}-V_{1(2)}\\
T_{N}-T_{1(2)}%
\end{array}
\right)  , \label{iq2}%
\end{eqnarray}
where $\tilde{V}_{1(2)}=(V_{N}+V_{1(2)})/2$ and $\tilde{T}_{1(2)}%
=(T_{N}+T_{1(2)})/2$. For reservoir temperatures $T_{1}=T_{2}=T$ and 
in the absence of the thermopowers $S_{1(2)}=0$, or when Joule heating dominates,
particle and energy current conservation requires $(G_{1}+G_{2})V_{N}=G_{1}V_{1}+G_{2}%
V_{2}$ and
\begin{equation}
T_{N}^{2}=T^{2}+\frac{(\Delta V)^{2}}{\mathcal{L}_{0}}\frac{G_{1}G_{2}}%
{(G_{1}+G_{2})^{2}}, \label{TN3}%
\end{equation}
so that we recover the result for the maximum amplitude of the electron temperature
profile in the middle of a diffusive bulk wire (with the conductance $G_1=G_2=G$) 
due to heating by inelastic electron-electron collisions.\cite{Nagaev:prb95,Pothier:prl97} By
taking into account the thermopowers of the junctions, but in the limit
$S^{2}\ll\mathcal{L}_{0}$, the change in the electron temperature in the
island is found as
\begin{eqnarray}
&& \Delta T_{N}=\frac{(\Pi_{1}-\Pi_{2})I}{\mathcal{L}_{0}T(G_{1}+G_{2})}%
+\frac{\bar{R}I^{2}}{2\mathcal{L}_{0}T(G_{1}+G_{2})}\nonumber\\
&& +\mathcal{O}\left(
\left(  \frac{\Delta T}{T}\right)  ^{2},\frac{\Delta T}{T}\frac{(\Delta V)^{2}}{\mathcal{L}_{0}T^{2}}
,\frac{S(\Delta V)^{3}}{\mathcal{L}_{0}^{2}T^{3}},\cdots\right)  . \label{TN2}%
\end{eqnarray}
in which the electric current Eq. (\ref{electric}) is driven by a voltage
and/or temperature bias. Equations (\ref{TN3}) and
(\ref{TN2}) yield the same result for a small temperature increase due to Joule heating  
when $\Pi_{1}=\Pi_{2}$ and $T_{1}=T_{2}$.

Following Fukushima \textit{et al.}\cite{Fukushima:jjap05,Fukushima:ieeem05}
we use Eq. (\ref{TN2}) to derive an expression for the Peltier coefficient in
terms of a critical current $I_{p}$ at which heating and cooling cancel each
other. Assuming that the resistance scales linearly with the electron
temperature in the node, $\Delta T_{N}=0$ leads to
\begin{equation}
(\Pi_{2}-\Pi_{1})I_{p}=\frac{\bar{R}I_{p}^{2}}{2}.\label{Ip}%
\end{equation}
The factor 1/2 on the right hand side implies that the current heating
is only half as large as considered by Refs. \onlinecite{Fukushima:jjap05} and \onlinecite{Fukushima:ieeem05}, 
whereas the \textquotedblleft cooling power\textquotedblright\ 
$\bar{R}I_{p}=2(\Pi_{2}-\Pi_{1})$ (in units of mV) is twice as large.
This discrepancy can be explained as follows. In our
model, the energy is dissipated in the nodes and reservoirs of the device, not
at the sharp interfaces. We monitor the temperature change in the normal metal
node which is assumed to be effectively thermalized. Half of the generated
heat is dissipated in the reservoirs that by definition do not contribute to
the resistance change. The expressions of Fukushima \textit{et al.}%
\cite{Fukushima:jjap05,Fukushima:ieeem05} can be recovered by
treating the highly resistive junctions in their samples as bulk material in
which heat is generated and contributes to its temperature and resistance rise
(see Section \ref{exp}).

\section{ Peltier and Seebeck effects in the presence of a single ferromagnetic
element}

The thermoelectric transport Eq. (\ref{IQ}) can be generalized to include the
spin degree of freedom. For spin-dependent thermoelectric transport through an
$F|N$ interface the spin-polarized electric charge and heat currents
read
\begin{equation}
\left(
\begin{array}
[c]{c}%
I^{\uparrow(\downarrow)}\\
\dot{Q}^{\uparrow(\downarrow)}%
\end{array}
\right)  =G^{\uparrow(\downarrow)}\left(
\begin{array}
[c]{cc}%
1 & S^{\uparrow(\downarrow)}\\
-S^{\uparrow(\downarrow)}T & -\mathcal{L}_{0}T
\end{array}
\right)  \left(
\begin{array}
[c]{c}%
V_{N}^{\uparrow(\downarrow)}-V_{F}^{\uparrow(\downarrow)}\\
T_{N}^{\uparrow(\downarrow)}-T_{F}^{\uparrow(\downarrow)}%
\end{array}
\right)  , \label{pars}%
\end{equation}
where the spin-dependence of the conductance $G^{\uparrow(\downarrow)}$,
thermopower $S^{\uparrow(\downarrow)}=-e\mathcal{L}_{0}T\partial_{\epsilon}\ln
G^{\uparrow(\downarrow)}|_{\epsilon_{F}}$, heat current, and temperature is
expressed by the superscript $\uparrow(\downarrow)$ for majority (minority)
spin electrons. $V_{s}=V^{\uparrow}-V^{\downarrow}\ $ is the \emph{particle
spin accumulation}. Referring to the discussion below we conjecture the
existence of a \emph{heat spin accumulation }$T_{s}=T^{\uparrow}%
-T^{\downarrow}$, \textit{i.e}., a temperature imbalance for majority and minority electrons, 
when thermalization is weak. We also define the total
thermopower $S=-(\Delta V/\Delta T)_{I=0}$ of an interface between a normal
metal and a ferromagnet as
\begin{equation}
S=-e\mathcal{L}_{0}T\left(  \frac{\partial_{\epsilon}G}{G}\right)
_{\epsilon_{F}}=\left(  \frac{G^{\uparrow}S^{\uparrow}+G^{\downarrow
}S^{\downarrow}}{G^{\uparrow}+G^{\downarrow}}\right)  _{\epsilon_{F}}.
\label{SM}%
\end{equation}
This thermopower is observable when the interface is part of a (hetero)
Sharvin point contact in direct contact with large reservoirs 
that prevent build-up of a spin accumulation. In
a diffusive environment, however, the local spin accumulation should be taken
into account, as described in the following. The spin-polarization of the
interface thermopower is defined as
\begin{equation}
P_{S}\equiv\frac{S^{\uparrow}-S^{\downarrow}}{S^{\uparrow}+S^{\downarrow}%
}=\frac{P^{\prime}-P}{1-P^{\prime}P}%
\end{equation}
where $P$ and $P^{\prime}$ are the polarizations of the conductance
$G^{\uparrow(\downarrow)}$ and its energy derivative $\partial_{\epsilon
}G^{\uparrow(\downarrow)}$ respectively, both at the Fermi energy. Whereas
$\left\vert P\right\vert <1,$ $\left\vert P^{\prime}\right\vert \gg1$ when
$\partial_{\epsilon}G^{\uparrow}$ approaches $-\partial_{\epsilon
}G^{\downarrow}$. $P_{S}$ is also in principle unbounded. Using 
\begin{equation}
-e\mathcal{L}_{0}T\left(  \frac{\partial P}{\partial\epsilon}\right)
_{\epsilon_{F}}=S(P^{\prime}-P)_{\epsilon_{F}}, \label{P-energy}%
\end{equation}
it follows that $P_{S}\neq0$ when the conductance polarization is energy
dependent. Spin-polarization of the thermopower of ferromagnetic
materials\cite{Cadeville:jpf71,Pirauxa:jmmm92,Shi:prb96} has been invoked to,
\textit{e.g}., explain the giant magneto-thermoelectric effect of magnetic
multilayers.\cite{Gravier:jmmm04} For a few combinations of materials, the
interface thermopower and its spin polarization are known from first
principles calculations.\cite{Hatami:prl07,Zhang:prb09}

Consider now an $F_{1}|N|N_{2}$ pillar with one magnetic contact. Conservation
of charge, spin and energy currents implies the Kirchhoff rules $I_{1\alpha
}+I_{2\alpha}=0$ and $\sum_{\alpha}\dot{Q}_{1}^{\alpha}+\dot{Q}_{2}^{\alpha
}=0,$ where $\alpha =\uparrow\left(  \downarrow\right)  $. 
The individual spin currents are separately conserved since we disregard spin-flip
scattering in the normal metal spacer when the length of the metal does not
exceed its spin diffusion length. In contrast, the heat spin accumulation,
\textit{i.e.} the temperature difference between the two spin species on the
central island, is assumed to vanish by strong inelastic scattering, 
which is likely for temperatures which
are not too low and/or metals which are not too clean.
In this regime the electron temperature on the island becomes%
\begin{equation}
T_{N}=T_{0}+\frac{(\gamma_{1}S_{1}-S_{2})I}{\mathcal{L}_{0}(G_{1}+G_{2})}.
\label{FTN}%
\end{equation}
We may call
\begin{equation}
\gamma_{1}=\frac{(1-P^{\prime}P)+G_{2}/G_{1}}{(1-P^{2})+G_{2}/G_{1}}.
\end{equation}
a "spin-entropy factor", because it reflects the spin-polarization of the
entropy flow per unit of the electric current (the thermopower), 
$S^{\uparrow}\neq S^{\downarrow}$ ($P^{\prime}\neq P$).  
In the limit $S_{\uparrow
(\downarrow)}^{2}\ll L_{0}$ (and therefore $\kappa^{\uparrow(\downarrow
)}\approx L_{0}TG^{\uparrow(\downarrow)}$) the temperature $T_{0}$ (Eq.
(\ref{T0})) is not affected by the magnetism. When $P^{\prime}\neq P$ 
the Peltier cooling (heating) does not vanish even when $S_{1}%
=S_{2}$. The thermopower spin polarization, $P_{S}$, can enhance or
suppress the Peltier effect depending on the spin polarization $P$ and the
relative amplitude of the conductances $G_{2}/G_{1}$. $\gamma_{1}$ can become
large when $\left\vert P^{\prime}\right\vert \gg1$ (an example is $P^{\prime}$
at a disordered Cr$|$Fe interface, see Table. I).

The total thermopower can be expressed in terms of the properties of its constituent elements,
in the limit $S^{\uparrow(\downarrow)}\ll \sqrt{\mathcal{L}_{0}}$ 
and for strongly thermalized electrons, as
\begin{equation}
\frac{\bar{S}}{\bar{G}}=\frac{\gamma_{1}S_{1}}{G_{1}}+\frac{S_{2}}{G_{2}}.
\label{S-FNN}%
\end{equation}
Therefore, when a spin accumulation is excited in the proximate normal metal, 
the magnetic junction contributes to the thermopower not by 
the Seebeck coefficient of the point contact $S_{1},$ but
by the product with the spin-entropy factor $\gamma_{1}S_{1}$.

\section{Magneto-Peltier and magnetothermopower in spin valves}

We proceed to the study of thermoelectric effects in asymmetric $F_{1}\left(
\mathbf{m_{1}}\right)  |N|F_{2}\left(  \mathbf{m_{2}}\right)  $ spin valves
(see Fig.~\ref{fig1}) for arbitrary relative orientations of the
magnetizations, $\mathbf{m_{1}}\cdot\mathbf{m_{2}}=\cos\theta$. The electron
distributions in the nodes and reservoirs are now $2\times2$ matrices in spin
space that can be expanded into scalar and vector components $\hat{f}%
^{F(N)}=f_{c}^{F(N)}\hat{1}+\hat{\mbox{\boldmath$\sigma$}}\cdot\mathbf{s}%
^{F(N)}f_{s}^{F(N)}$, where $\hat{\mbox{\boldmath$\sigma$}}$ is the vector of
Pauli matrices and $\hat{1}$ the $2\times2$ unit matrix. The unit vector of
the spin quantization axis $\mathbf{s}^{F}$ is parallel to the magnetization
of the ferromagnet, whereas $\mathbf{s}^{N}$ can point in any direction. In
linear response, the $2\times2$ spectral current in spin space across a
ferromagnet-normal metal junction at energy $\epsilon$ in the absence of
spin-flip and inelastic interface scattering is given as a spectral
Landauer-B\"{u}ttiker-like
expression\cite{Brataas:prp06,Brataas:prl00,Brataas:epjb01}
\begin{equation}
\hat{\imath}_{N|F}\left(  \epsilon\right)  =\sum_{\alpha\beta}G^{\alpha\beta
}\left(  \epsilon\right)  \hat{u}^{\alpha}[\hat{f}^{F}\left(  \epsilon\right)
-\hat{f}^{N}\left(  \epsilon\right)  ]\hat{u}^{\beta}, 
\label{i-sv}
\end{equation}
where $\hat{u}^{\uparrow(\downarrow)}=(\hat{1}\pm\hat
{\mbox{\boldmath$\sigma$}}\mathbf{\cdot m})/2$ are projection matrices in
which the unit vector $\mathbf{m}\equiv\mathbf{s}^{F}$ denotes the magnetization direction of the
ferromagnet. The conductance tensor elements read $G^{\alpha\beta}\left(
\epsilon\right)  =\left(  e^{2}/h\right)  \sum_{nm}[\delta_{mn}-r_{nm}%
^{\alpha}\left(  \epsilon\right)  (r_{nm}^{\beta}\left(  \epsilon\right)
)^{\ast}]$ in terms of the energy-dependent reflection coefficients
$r_{nm}^{\alpha}(\epsilon)$ for majority and minority spins at the $N|F$
interface. Its diagonal elements are the conventional spin-dependent
conductances that govern, \textit{e.g}., the giant magnetoresistance, whereas
the complex non-diagonal elements, the so-called spin-mxing conductances,
parameterize the transverse spin currents that are absorbed by the ferromagnet
and give rise to torques on the magnetization. The total charge-spin and heat
matrix currents are defined as $\hat{I}=\int d\epsilon\;\hat{\imath}%
(\epsilon)$ and $e\hat{\dot{Q}}=\int d\epsilon(\epsilon-\mu)\hat{\imath
}(\epsilon)=e\hat{I}^{\epsilon}-\mu\hat{I},$ respectively, where $\mu$ is the
equilibrium chemical potential and $\hat{I}^{\epsilon}$ the energy current. In
the following we assume that both spin components $f^{\uparrow(\downarrow
)}=f_{c}\pm f_{s}$ of the diagonalized matrix distribution functions $\hat
{f}^{F(N)}$ may be described by thermal-equilibrium Fermi-Dirac distribution
functions with spin-dependent chemical potentials and temperatures. The
Sommerfeld expansion can then be employed to derive expressions for the
transport currents as a function of applied voltage or temperature gradients
in terms of the conductance tensor $G^{\alpha\beta}$ and its energy derivative
$G_{\epsilon}^{\alpha\beta}\ $at the Fermi energy.\cite{Hatami:prl07} The
total charge, spin and heat currents read $I_{c}=\mathrm{Tr}[\hat{I}],$
$\mathbf{I}_{s}=\mathrm{Tr}[\mbox{\boldmath $\sigma $}\hat{I}]$ and $\dot
{Q}_{c}=\mathrm{Tr}[{\hat{\dot{Q}}}],$ respectively, where the trace is over
spin indices.

The charge and energy conservation laws read $I_{c1}+I_{c2}=0$ and $\dot
{Q}_{c1}+\dot{Q}_{c2}=0$. Moreover, in the absence of spin-flip scattering in
the normal node, the total spin angular momentum current is conserved as well,
\textit{i.e.}, $\mathbf{I}_{s1}+\mathbf{I}_{s2}=0$. These Kirchhoff Laws close
the system of transport equations in the strongly thermalized regime. In what
we call the \emph{weakly thermalized regime}, the distributions for each spin
species are thermalized separately, but the energy exchange between the spin
subsystems is disregarded, which is a realistic scenario at low
temperatures.\cite{Hatami:Unp} In this limit a spin temperature vector on the
central island exists, $\mathbf{T}_{s}\neq0,$ and we require $\mathbf{\dot{Q}%
}_{s1}+\mathbf{\dot{Q}}_{s2}\approx0$ where $\mathbf{\dot{Q}}_{s}%
=\mathrm{Tr}[\mbox{\boldmath$\sigma $}\hat{\dot{Q}}]$, which means that energy
is conserved for each spin channel separately. 
It is worth while to compare the thermoelectric transport properties such as the total
conductance and thermopower of the spin-valve structure in the different interacting regimes.

In the strongly thermalized regime for a symmetric spin-valve, $G_{1}^{\alpha\beta}%
=G_{2}^{\alpha\beta}$ and $G_{\epsilon1}^{\alpha\beta}=G_{\epsilon2}%
^{\alpha\beta},$ the temperature of the normal metal island
$T_{N}=T_{0}$ is not affected by electric current. 
The total electric current reads\cite{Hatami:prl07}
\begin{equation}
I_{c}=\frac{G}{2}\left(  \Delta V+S\Delta T\right)  -\frac{PG}{2}\frac
{\tan^{2}\theta/2}{\eta_{R}+\tan^{2}\theta/2}\left(  P\Delta V+P^{\prime
}S\Delta T\right)  .\label{Ic-sv}%
\end{equation}
Here $\eta_{R}=2\operatorname{Re}G^{\uparrow\downarrow}/G>0,$ where
$G^{\uparrow\downarrow}$ is the complex spin mixing
conductance.\cite{Brataas:prp06} For most metallic contacts 
$\eta_{I}=2\operatorname{Im}G^{\uparrow\downarrow}/G,$ is
small ($\eta_{I}\ll\eta_{R}$) \cite{Zwierzycki:prb05} and is disregarded in the analytical results.
However, since $\eta_{I}$ is not small for $\text{Cr}|\text{Fe}$ and $\text{Cr}|\text{Co}$ 
interfaces, see Table I, it is included in the numerical results for these junctions.
The angular magneto-resistance for $\Delta T=0$ as measured by Urazhdin
\textit{et al}.\cite{Urazhdin:prb05} is well described by circuit
theory.\cite{Kovalev:prb06} The thermoelectric transport properties of the
spin-valve structure differ significantly in the different interacting
regimes. In the Sommerfeld approximation, the spin-mixing thermopower $S^{\uparrow\downarrow}%
\equiv-e\mathcal{L}_{0}TG_{\epsilon}^{\uparrow\downarrow}/G^{\uparrow
\downarrow}$ and the dimensionless mixing parameter $\eta_{R(I)}^{\prime}%
\equiv2\operatorname{Re}(\operatorname{Im})G^{\uparrow\downarrow}_{\epsilon}%
/G_{\epsilon}$ enter expressions for the electric currents only when ${\bf T}_s^{N}\neq 0$,
i.e. in the weakly thermalized regime.\cite{Hatami:prl07}

\begin{table*}[ptb]
\begin{center}
\begin{tabular}
[c]{lcrrrrrrrr}\hline\hline
 & $G(10^{15}\operatorname{\Omega}^{-1}\operatorname{m}^{-2})$ & $S/T$%
($\operatorname{nV}/\operatorname{K}^{2}$) & $P(\%)$ & $P^{\prime}(\%)$
& $P_{S}(\%)$ & $\eta_{R}$ & $\eta_{I}$ & $\eta_{R}^{\prime}$ &
$\eta^{\prime}_{I}$\\\hline
Cu$|$Co(001)  & 4.43 & -13 &  75 &   72 &    -8 & 0.50 & -0.036 &  0.03 &  0.00 \\
Cu$|$Co(001)* & 4.29 & -34 &  74 &   89 &    43 & 0.49 & -0.054 &  0.06 & -0.01 \\\hline
Cu$|$Co(110)  & 3.42 & -10 &  69 &    6 &   -66 & 0.67 & -0.082 & -0.32 &  0.44 \\
Cu$|$Co(110)* & 3.52 & -13 &  64 &   85 &    45 & 0.63 & -0.077 &  0.07 & -0.05 \\\hline
Cu$|$Co(111)  & 3.69 & -15 &  60 &   56 &    -6 & 0.53 & -0.006 &  0.13 &  0.48 \\
Cu$|$Co(111)* & 3.42 & -15 &  68 &   77 &    17 & 0.64 & -0.073 &  0.13 & -0.05 \\\hline
Cr$|$Au(001)  & 0.36 &   7 &   0 &    0 &     0 &  $-$ &    $-$ &   $-$ &   $-$ \\
Cr$|$Au(001)* & 0.67 &   0 &   0 &    0 &     0 &  $-$ &    $-$ &   $-$ &   $-$ \\\hline
Cr$|$Fe(001)  & 0.88 &  22 & -74 &  -40 &    48 & 4.23 &   1.38 & -4.27 & -1.38 \\
Cr$|$Fe(001)* & 0.94 &   7 & -53 & -190 & -9500 & 3.25 &   0.43 & -0.48 & 9.39 \\\hline
Cr$|$Co(001)  & 0.56 &  62 & -62 & -111 &  -160 & 3.03 &  -0.59 & -2.86 & -3.46 \\
Cr$|$Co(001)* & 0.71 &  23 & -23 &  -95 &   -92 & 2.92 &  -1.79 & -0.86 & 3.21 \\\hline\hline
\end{tabular}
\end{center}
\caption{Thermoelectric interface parameters calculated at the Fermi
energy for a number of almost lattice-matched interfaces including Schep's 
drift correction.\cite{Schep:prb97} 
With the exception of the Cr$|$Co interface for which Co is assumed to be bcc,
Cr and Fe are bcc while Cu, Co and Au are assumed to be fcc.
The asterisk * indicates a dirty interface modeled in a
10x10 lateral supercell with two layers of 50\%-50\% alloy.}%
\label{coefficient}%
\end{table*}

In the presence of a temperature bias $\Delta T$ with an open electric circuit
($I_{c}=0)$, the induced thermoelectric voltage $\Delta V$ is described by the angular
magnetothermopower (MTP) $\bar{S}(\theta)=-\left(  \Delta V/\Delta T\right)
_{I_{c}=0}$ which in the strongly thermalized regime reads \cite{Hatami:prl07} 
\begin{equation}
\frac{\bar{S}(\theta)}{S}=\frac{\eta_{R}+(1-PP^{\prime})\tan^{2}\theta/2}%
{\eta_{R}+(1-P^{2})\tan^{2}\theta/2},\ \label{MT}%
\end{equation}%
\begin{equation}
MTP\equiv\frac{\bar{S}(\pi)-\bar{S}(0)}{S}=\frac{-PP_{S}}{1+PP_{S}}.\label{MTP}%
\end{equation}
The MTP is finite when the interface thermopower is spin polarized, $P_{S}\neq0$. 
When $PP^{\prime}>1$ (which also requires $P_{S}<0 $) one finds an angle 
$\theta_0=2\tan^{-1}\sqrt{ \eta_{R}/(PP^{\prime}-1)}$ where the thermoelectric power
can change sign, \textit{i.e.} a spin-valve with non-collinear magnetic configuration 
can display a transition from electron-like to hole-like transport. 
In a spin-valve the thermally induced charge-currents that enter the normal node from both sides 
can be made to cancel such that the net thermoelectric voltage vanishes, ${\bar S}=0$.
Note that not only the individual spin currents but also the charge current depend on the effective spin polarizations. 
The MTP vanishes in the half-metallic limit $P=P^{\prime}=\pm 1$.

In the weakly thermalized regime on the other hand, the magnetothermopower
Eq. (\ref{MTP}) is twice as large, whereas the magnetoresistance and $\bar
{S}(0)$ do not change (in the limit $S^{2}\ll\mathcal{L}_{0}$). $\bar{S}(\pi)$
is enhanced in this case because the heat spin accumulation facilitates the
thermoelectric voltage build-up. In this regime an angular magnetothermopower 
is found even when $P_{S}=0$ provided that 
$\eta_{R}^{\prime}\equiv2\operatorname{Re}%
\partial_{\epsilon}G^{\uparrow\downarrow}/\partial_{\epsilon}G\neq\eta_{R}$,
, which is destroyed by full thermalization.

The Onsager-Kelvin relation between the total Seebeck 
and Peltier coefficients in spin valves, 
\textit{i.e.} ${\bar \Pi}(\theta)={\bar S}(\theta) T$ 
in which ${\bar \Pi}\equiv\left( -{\dot Q}_{c}/I_{c} \right)_{\Delta T=0}$, 
is found to hold in both thermalization regimes.

For a quantitative analysis we need to know more about the
thermoelectric parameters for interfaces. In the absence of
experimental estimates, we calculated the parameters from
first-principles within the framework of density functional theory for
a number of interfaces which figure prominently in the field of
magnetoelectronics.\cite{Hatami:prl07} The values are given in
Table~\ref{coefficient}. For an A$|$B interface, the calculation
proceeds as follows.\cite{Xia:prb06,Zwierzycki:pssb08} Self-consistent
density functional theory calculations are first performed separately
for bulk A and B materials. These calculations yield bulk charge- and
spin- densities and potentials, and the corresponding Fermi energies. A
self-consistent interface calculation is next performed subject to the
potentials (and densities) far from the interface being equal to their
bulk values, up to a constant which is adjusted so as to equalize the
Fermi energies.\cite{Turek:97} The interface breaks the lattice
periodicity perpendicular to the interface leaving only two-dimensional
periodicity parallel to the interface that is characterized by the
two-dimensional Bloch vector $\mathbf{k}_{\parallel}$. The electronic
structure of the localized perturbation formed by the interface is
handled using a Green's function method, a so-called ``Surface Green's
Function''. The rank of the matrix of the perturbation is made finite
and minimized by making use of the translational symmetry parallel to
the interface and using a maximally-localized basis of tight-binding
(TB) muffin-tin orbitals (MTOs).\cite{Andersen:prl84,Andersen:prb86} To
calculate the scattering matrix $\mathcal{S}$
\begin{equation}
\mathcal{S}(\mathbf{k}_{\parallel},\epsilon) \equiv \left(
\begin{array}
[c]{cc}%
r(\mathbf{k}_{\parallel},\epsilon) & t^{\prime}(\mathbf{k}_{\parallel
},\epsilon)\\
t(\mathbf{k}_{\parallel},\epsilon) & r^{\prime}(\mathbf{k}_{\parallel
},\epsilon)
\end{array}
\right)  ,\label{scatt}%
\end{equation}
at real energies (at or close to the Fermi energy in the context of
transport), we use a wave-function-matching scheme due to
Ando\cite{Ando:prb91} which involves the calculation of individual
scattering states far from the interface. The rank of the reflection and transmission matrices
$r,r^{\prime },t,t^{\prime}$ is determined by the number of Bloch
states at a given energy $\epsilon$ and transverse wave-vector
$\mathbf{k}_{\parallel}$. The minimal TB-MTO basis is very efficient
and makes it possible to model incommensurate lattices and various
types of disorder using large lateral
supercells.\cite{Xia:prb01,Xia:prb06,Zwierzycki:pssb08} Substitutional
disorder where one or more layers of atoms form an alloy is
conveniently treated by calculating the potentials self-consistently
using a layer version\cite{Turek:97} of the coherent potential
approximation\cite{Soven:pr67} and then distributing at random the site
potentials in lateral supercells subject to maintenance of the
appropriate layer concentrations.\cite{Xia:prb06} The mixing
conductance is most easily calculated in terms of the reflection
matrices.\cite{Xia:prb02,Zwierzycki:prb05} We consider here interfaces
in diffuse metallic systems which implies that we have to use a
generalization of the Schep correction\cite{Schep:prb97,Brataas:prp06}
by replacing the bare $G^{\alpha\beta}(\epsilon)^{-1}$ with
$G^{\alpha\beta}(\epsilon)^{-1}- \left[
G_{\mathrm{N}}(\epsilon)^{-1} +
\delta_{\alpha\beta}G^{\alpha}_{\mathrm{F}}(\epsilon)^{-1}\right] $,
where
$G_{\mathrm{N}}(\epsilon)=e^2 M_{\mathrm{N}}(\epsilon)/h$ and 
$G^{\alpha}_{\mathrm{F}}(\epsilon)=e^2 M^{\alpha}_{\mathrm{F}}(\epsilon)/h$ are the
single-spin Sharvin conductances of the normal and ferromagnetic metals
forming the interface. The thermopower and other generalized
thermoelectric parameters are determined by numerically 
differentiating the scattering matrix calculated as a function of the energy.
Details of the numerical procedures will be given in a separate paper.\cite{Zhang:prb09}
\begin{figure}[t]
\resizebox{\columnwidth}{!} {\includegraphics{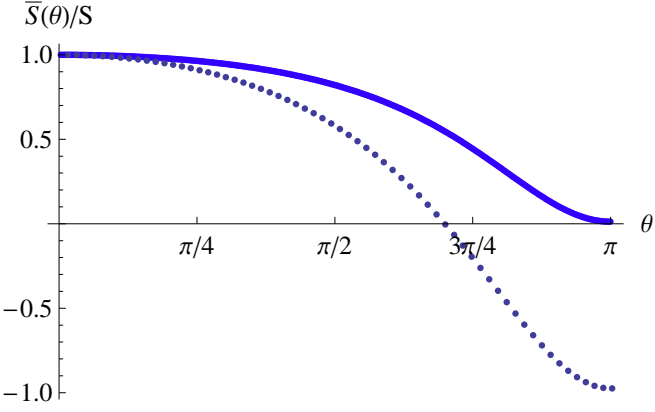}}
\caption{Magnetothermoelectric power as a function of relative angle between
the magnetizations in a $\mathrm{Fe}\mathrm{|Cr|Fe}$ (001) spin valve
with dirty interfaces. The MTP is significantly different for
the strong (full line) and weak (dotted line) thermalization in the normal
metal spacer.}%
\label{fig2}%
\end{figure}

\begin{figure}[t]
\resizebox{\columnwidth}{!} {\includegraphics{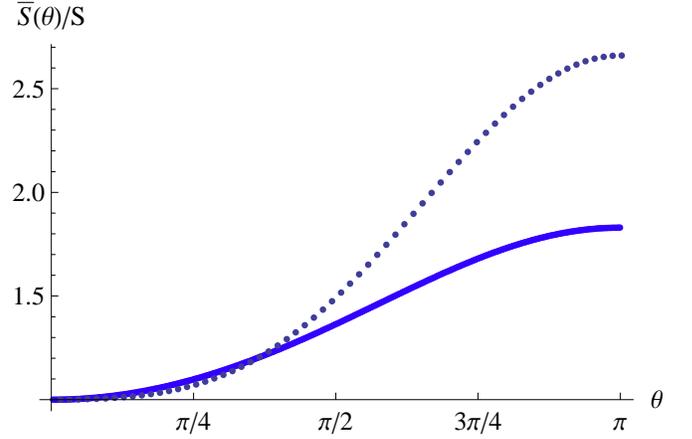}}
\caption{Same as Fig. \ref{fig2} but for a $\mathrm{Co|Cu|Co}$ (110) spin valve with
clean interfaces.}%
\label{fig3}%
\end{figure}
 
We use the data in Table~I to compute the angular dependence of the
thermoelectric properties of a few spin valves for illustrative
purposes. Spin-flip scattering is disregarded here but will be
discussed in later sections.

We plot the angular MTP for a Fe$|$Cr$|$Fe (001) spin valve with
dirty interfaces in Fig. \ref{fig2} and for a Co$|$Cu$|$Co (110) spin valve with clean interfaces
in Fig. \ref{fig3}. The angular MTP is enhanced in the weakly
thermalized regime (shown by dotted lines) by up to a factor of two for
the antiparallel configuration.
Depending on $P$ and $P_{S}$, the MTP can be of any sign, see Eq. (\ref{MTP}).

We now turn to the Peltier effect of asymmetric spin valves. In the strongly
thermalized regime and for $P_{S_{1}}=P_{S_{2}}=0,$ the Peltier cooling retains the simple
form for normal metal structures, Eq. (\ref{TN1}), whereas the total charge
current is a complicated function of the magnetic configuration of the system.
A magneto-Peltier effect (MPE), \textit{i.e}., a dependence of the cooling
power on the magnetic configuration, is found when the thermopower is spin dependent. 
For a voltage-biased spin valve with
thermal asymmetry $S_{1}\neq S_{2}$ and $P_{1}^{\prime}\neq P_{2}^{\prime},$
but  $G_{1}=G_{2}=G$, $P_{1}=P_{2}=P,$ and $S^{\uparrow(\downarrow)}_{i}%
\ll \sqrt{\mathcal{L}_{0}}$, we find for the temperature change of the normal metal
spacer
\begin{equation}
\Delta T_{N}=\frac{(\gamma^{\text{MP}}_{1}S_{1}-\gamma^{\text{MP}}_{2}S_{2})I_{c}}{2\mathcal{L}_{0}%
G},\label{DTF}%
\end{equation}
where the spin-entropy factors
\begin{equation}
\gamma_{1(2)}^{\text{MP}} \left( \theta \right) =\frac{(1-P_{1(2)}^{\prime}P)\tan
^{2}\theta/2+\eta_{R}}{(1-P^{2})\tan^{2}\theta/2+\eta_{R}}%
\end{equation}
depend now on the magnetic configuration ($P^{\prime}_{1(2)}\neq P$). 
$\gamma_{1(2)}^{\text{MP}}=1$ and $\left(1-P_{1(2)}^{\prime}P\right)  /\left(  1-P^{2}\right) ,$ 
respectively, for parallel and antiparallel configurations. The MPE should therefore be
observable in R \textit{vs.} I curves of spin valves during current-induced
magnetization reversal. According to Eq. (\ref{Ip}) with 
$\bar{R}(I=I_p)=\bar{R}(I=0)\equiv R_{0},$ the temperature change $\Delta T_{N}=T_{N}%
(\theta)-T_{0}$ corresponds to the cooling-power $R_{0}I_{p}\approx
2\mathcal{L}_{0}(G_{1}+G_{2})T\Delta T_{N}/I_{p}$ such that
\begin{eqnarray}
&&R_{0}\left[  I_{p}(\pi)-I_{p}(0)\right]  =\nonumber\\
&&4\frac{G_{1}P_{2}(P_{1}^{\prime
}-P_{1})\Pi_{1}-G_{2}P_{1}(P_{2}^{\prime}-P_{2})\Pi_{2}}{G_{1}(1-P_{1}%
^{2})+G_{2}(1-P_{2}^{2})}\label{MPsignal}%
\end{eqnarray}
This signal contains unique information on the spin-polarization of the thermopower.
When thermalization is weak a MPE arises even when $P_{S_1}=P_{S_2}=0$ 
($P^{\prime}_{1(2)}=P_{1(2)}$).
A sign change in the cooling-power is also expected. In the strongly 
thermalized regime this arises from different angular dependences of 
the spin-entropy factors. When the effective thermopowers are equal, 
$\gamma^{\text{MP}}_{1}(\theta_0)S_{1}= \gamma^{\text{MP}}_{2}(\theta_0)S_{2}$, 
no Peltier cooling is expected, $R_0 I_p(\theta_0)=0$.

\begin{figure}[t]
\resizebox{\columnwidth}{!} {\includegraphics{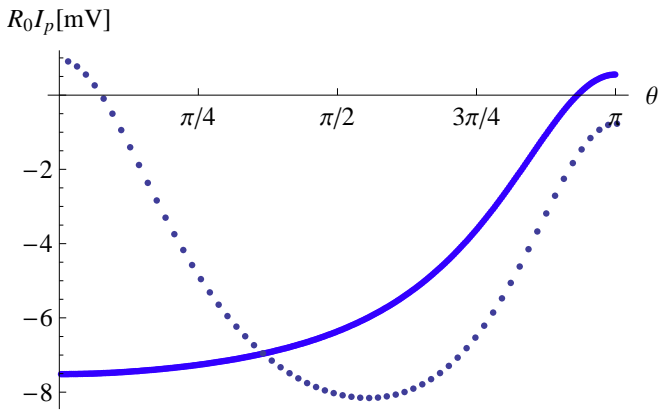}}
\caption{Magneto-Peltier cooling as a function of the angle between the
magnetizations in the hypothetical bcc $\mathrm{Co|Cr|Fe}$ (001) spin valve. 
$R_{0} I_{p}(\theta)$ is the cooling power which compensates local Joule heating. 
The signal $R_{0}[I_{p}(\pi)-I_{p}(0)]$ is significantly different for strong (full line)
and weak (dotted line) thermalization in the normal metal spacer.}%
\label{fig4}%
\end{figure}

\begin{figure}[t]
\resizebox{\columnwidth}{!} {\includegraphics{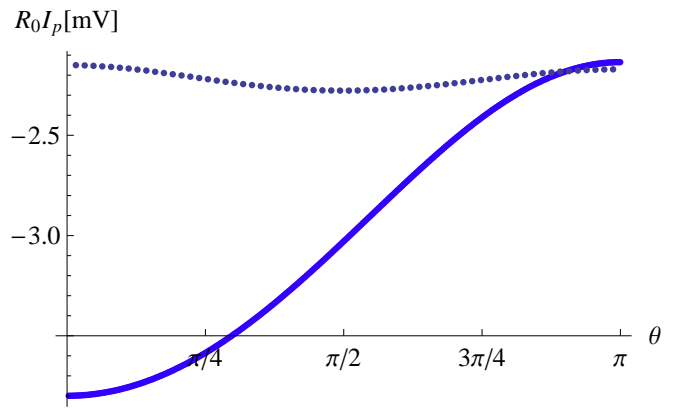}}
\caption{As Fig. \ref{fig4} but for an asymmetric $\mathrm{Co}|\mathrm{Cu}%
|\mathrm{Co}$ (001) spin valve with one clean and one disordered interface.}%
\label{fig5}%
\end{figure}

In Figs.~\ref{fig4} and \ref{fig5}, we illustrate the magneto-Peltier
cooling by computing the angular dependent cooling-power $R_0 I_p(\theta)$, 
at room temperature, for a hypothetical bcc Co$|$Cr$|$Fe (001) spin-valve structure with
clean interfaces and for an asymmetric Co$|$Cu$|$Co (001)
with one ideal and one disordered interface, using parameters from Table~I. 
For comparison, we include the results in the
weakly thermalized regime, depicted in the figures by dotted lines. The
dependence of the Peltier cooling on the magnetic configuration 
observed in Fig. \ref{fig4} for the Co$|$Cr$|$Fe spin-valve is caused
by the relatively large values of the interface thermopower parameters
$P$ and $P^{\prime}$. The relatively weak magneto-Peltier signal,
$R_0[I_p(\pi)-I_p(0)]$ vanishes when $G_1 = G_2$ and $P_1=P_2$, for weakly
thermalized electrons (dotted line) is strongly enhanced by the
inter-spin energy exchange in the opposite limit. The MPE for the asymmetric
Co$|$Cu$|$Co (001) structure, Fig.~\ref{fig5}, displays similar effect but
with smaller amplitudes. For the strongly thermalized electron case,
the MPE caused by interface scattering, $R_{0}[I_{p}(\pi)-I_{p}(0)]$,
is of the order of $1-10 {\rm mV}$, which is smaller than the
experimental values\cite{Fukushima:jjap05,Fukushima:ieeem05} $R_0 I_p
\approx2 0-40 {\rm mV}$. For a typical current density of $10^8 {\rm
A/cm}^2$, we find a maximum temperature drop $\Delta T_N
(\theta)\approx 0.15 {\rm K}$ which is also too small to explain
experiments. Note that in the above we have assumed that interface
scattering is dominant. As explained in the next sections, we show that
these numbers are increased by including scattering in the bulk.

The magneto-Peltier effect vanishes for symmetric spin-valve structures.
Introducing the asymmetry $G_{1}\neq G_{2},$ we obtain an angle-dependent temperature
modulation
\begin{eqnarray}
&& \mathcal{L}_{0}\Delta T_{N}(\theta)=\nonumber\\
&& \frac{2\left(  G_{1}-G_{2}\right)  P\left(
P^{\prime}-P\right)  SI_{c}\sin^{2}\frac{\theta}{2}}{(G_{1}+G_{2})^{2}%
(1-P^{2})\eta_{R}+G_{1}G_{2}(1-P^{2}-\eta_{R})^{2}\sin^{2}\theta},\nonumber\\
\label{TNnc}%
\end{eqnarray}
even for equal thermopowers and spin polarizations at the interfaces.
Analytical expressions for the Peltier cooling in the weak thermalization
regime, in which the spin-mixing thermopower ($\eta_{R(I)}^{\prime}$ in Table I) 
becomes a relevant parameter, are much more complex. The computation
is straightforward, however, and is easily carried out when the necessity arises.

\section{Spin-conserving bulk impurity scattering}

In this section we discuss the contribution of bulk scattering for the case of
wires with constant cross section $A_{0},$ first for non-magnetic metals
and then for magnetic structures, both in the strongly thermalized limit.

A normal metal pillar is a heterostructure $\mathcal{R}_{A}|N_{A}%
(L_{A})|N\left(  L\right)  |N_{B}(L_{B})|\mathcal{R}_{B}$, where $N_{i}(L_{i})$ 
denotes a layer of material $i$ (=A, B) with thickness $L_{i}$ that can be 
larger than the elastic mean-free-path due to disorder scattering.
The length $L$ of the central island $N$ is so short that its (bulk) 
resistance can be disregarded. The external
reservoirs $\mathcal{R}_{A\left(  B\right)  }$ are in thermal equilibrium but
at different temperatures and/or voltages. The spreading resistance at an
abrupt opening can be accounted for by an effective length
parameter.\cite{Gravier:jpd06} The electron distribution functions in the
disordered metal wires follow from the diffusion equation in the bulk and are
connected at the interfaces by (quantum mechanical) boundary
conditions.\cite{Brataas:prp06} The conserved particle/heat currents can be
obtained from Eq. (\ref{IQ}) by replacing $-(V_{1(2)}-V_{N})$ by
$L_{A\left(  B\right)  }\nabla\mu_{A\left(  B\right)  }/e$ , $T_{1(2)}%
-T_{N}$ by $L_{A\left(  B\right)  }\nabla T_{A\left(  B\right)  }$ and
the interface conductance Eq. (\ref{G}) by the electric conductivity
$\sigma_{A\left(  B\right)  }=e^{2}\mathcal{N}_{A\left(  B\right)
}D_{A\left(  B\right)  }$, where $\mathcal{N}$ and $D$ are energy-dependent
densities of states and diffusion constants of the bulk materials,
respectively. Mott's formula,  $S_{A(B)}=-e\mathcal{L}_{0}T\partial_{\epsilon}\ln \sigma_{A(B)}|_{\epsilon_{F}}$,
holds for the diffusion thermopower which usually dominates at high temperatures.\cite{Gripshover:pr67,Bass:McGraw-Hill}
In linear response charge and energy current conservations imply 
$\nabla^{2}\mu=0$ and $\nabla^{2}T=0$. The chemical potential and
temperature depend linearly on position except for jumps at the 
contacts which are governed by the interface parameters. 
The local chemical potential and temperature are 
then found by the charge and energy current conservations at the boundaries.

As a function of the applied electric current we obtain the following expression for 
the temperature change $\Delta T_{N}$ on the normal metal island 
\begin{eqnarray}
&&\Delta T_{N}=\nonumber\\
&&\left(  \frac{S_{A}-S_{B}}{G_{A}G_{B}}+\frac{S_{1}-S_{2}}%
{G_{1}G_{2}}+\frac{S_{A}-S_{2}}{G_{A}G_{2}}+\frac{S_{1}-S_{B}}{G_{1}G_{B}%
}\right)  \frac{G_{tot}I}{\mathcal{L}_{0}},\nonumber\\
\label{bulk1}%
\end{eqnarray}
where $G_{A\left(  B\right)  }=\sigma_{A\left(  B\right)  }A_{0}/L_{A\left(
B\right)  }$ and $S_{A\left(  B\right)  }$ are the bulk (Drude) conductances and
thermopowers in the leads, $G_{tot}%
^{-1}=\sum_{i=1,2,A,B}G_{i}^{-1}$ is the total series conductance. The
interface contribution to the Peltier cooling disappears when $S_{1}=S_{2}$
and $G_{1}/G_{2}=G_{A}/G_{B}$. When $S_{A}=S_{B}$ and $\sigma_{A}=\sigma_{B}$, 
Peltier cooling is possible for different lengths of the normal leads
($L_{A}\neq L_{B}$).\cite{Gurevich:sst05} Eq. (\ref{bulk1}) can be simplified
by introducing the lumped conductances $G_{L}=G_{A}G_{1}/(G_{A}+G_{1})$ and
$G_{R}=G_{2}G_{B}/(G_{2}+G_{B})$ as well as the thermopowers $S_L$ and $S_R$
with $S_{L}/G_{L}=S_{A}/G_{A}+S_{1}/G_{1}$ and $S_{R}/G_{R}=S_{2}/G_{2}+S_{B}/G_{B}$ for the
left and right parts of the normal island. In terms of the new parameters we
find
\begin{equation}
\Delta T_{N}=\frac{(S_{L}-S_{R})I}{\mathcal{L}_{0}(G_{L}+G_{R})},
\end{equation}
which, as expected, has the same form as Eq. (\ref{TN1}) in the limit
$S^{2}\ll\mathcal{L}_{0}$.

Replacing the normal lead $N_{A}$ by a magnetic lead, say $F_{A}$, we find
that the thermoelectric cooling obeys Eq. (\ref{FTN}) after replacing the
interface conductances and thermopowers $G_{1(2)}$ and $S_{1(2)}$ by
$G_{L(R)}$ and $S_{L(R)}$, provided that the spin polarizations in bulk layers
and contacts are the same. A more complicated structure like $\text{Co}%
|\text{Au}|\text{Ti}|\text{Au}$ can be shown to be equivalent to an
F$|\mathrm{N}|\mathrm{N}_{2}$ pillar by a similar lumping of parameters.

We now turn to the MPE, \textit{i.e}. the dependence of the Peltier cooling on
the magnetic configuration of a spin-valve structure, in the presence of bulk
scattering. A simple analytical expression for the cooling-power 
(or the local Joule heating compensation current $I_p$) can be 
obtained when the spin polarizations of the bulk and interfaces are equal:%
\begin{equation}
R_{0}I_{p}=2(\gamma_{L}^{MP}\Pi_{L}-\gamma_{R}^{MP}\Pi_{R}),\label{SVP1}%
\end{equation}
where the spin-entropy factors $\gamma_{L(R)}^{MP}$ are
\begin{equation}
\gamma_{L}^{MP}=\frac{G_{L}\left[  1-P_{A}P_{B}-\left(  P_{A}-P_{B}\right)
P_{A}^{\prime}\right]  +G_{R}(1-P_{B}^{2})}{G_{L}(1-P_{A}^{2})+G_{R}%
(1-P_{B}^{2})}.\label{SVP2}%
\end{equation}
An expression for $\gamma_{R}^{MP}$ is obtained by interchanging the indices
$L\leftrightarrow R$ and $A\leftrightarrow B$. Eqs. (\ref{SVP1},\ref{SVP2})
reduce to Eq. (\ref{MPsignal}) when bulk scattering is disregarded.

A phonon (or magnon) thermal current can transfer momentum to the electrons in
the presence of inelastic scattering which in turn generates an additional 
electric field and modifies the thermopower. 
For normal (as well as ferromagnetic) metals
at sufficiently low temperatures a contribution of the phonon 
(and magnon) -drag effect may become significant.\cite{Blatt:76,Blatt:prl67}
The magnon-drag effect is likely to be suppressed strongly in heterostructures 
since magnons cannot escape the ferromagnets. Strong phonon scattering 
at interfaces will likewise reduce the phonon-drag effect in multilayers. 
A microscopic treatment of the phonon-drag effect in heterostructures 
is beyond the scope of the present paper, however. At elevated temperatures,
where the drag effect can be disregarded, Mott's formula holds approximately   
even in the presence of inelastic scattering.\cite{Jonson:prb80,Kontani:prb03}

\section{Spin-flip bulk impurity scattering}

\begin{figure}[t]
\resizebox{\columnwidth}{!} {\includegraphics{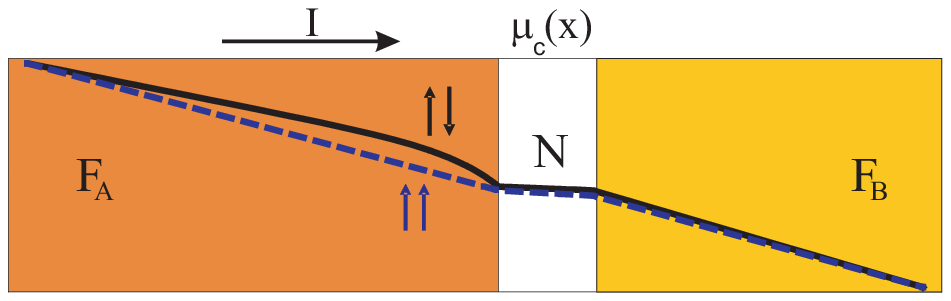}} \vspace{0.4 cm}
\resizebox{\columnwidth}{!} {\includegraphics{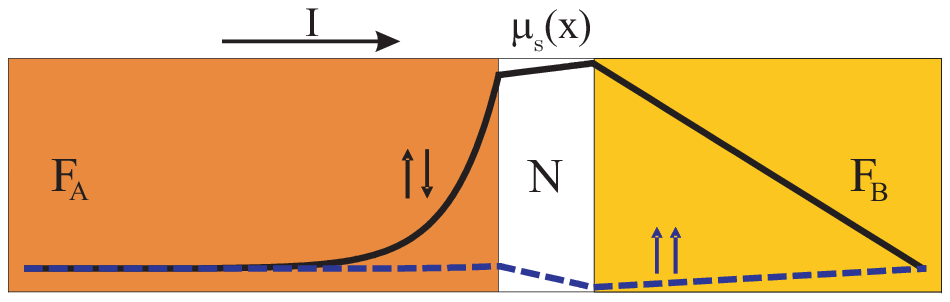}} 
\caption{Illustration of the local charge and spin chemical potentials in a
$\mathrm{F_{A}|N|F_{B}}$ spin valve biased by a voltage difference, for both
parallel ($\uparrow\uparrow$) and anti-parallel ($\uparrow\downarrow$)
alignments of the magnetizations. $\mathrm{F_{B}}(P=0.8,\lambda_{B}%
=0.1,S_{B}=-1\operatorname{\mu V}\mathrm{/}\mathrm{\operatorname{K}}$) has
been chosen to have weak spin-flip scattering and thermopower compared with
$\mathrm{F_{A}}(P=0.8,\lambda_{A}=10,S_{A}=-20\operatorname{\mu V}%
\mathrm{/}\mathrm{\operatorname{K}}$). The thin normal metal spacer is chosen
to be highly conductive ($\rho_{N}\ll\rho_{A(B)}=10\operatorname{\mu \Omega}%
\mathrm{\operatorname{cm}}$) and to have a thermopower equal to that of $\mathrm{F_{B}}$.
Note that spin accumulation is assumed to vanish at the two ends (at reservoirs).}%
\label{fig6}
\end{figure}

\begin{figure}[t]
\resizebox{\columnwidth}{!} {\includegraphics{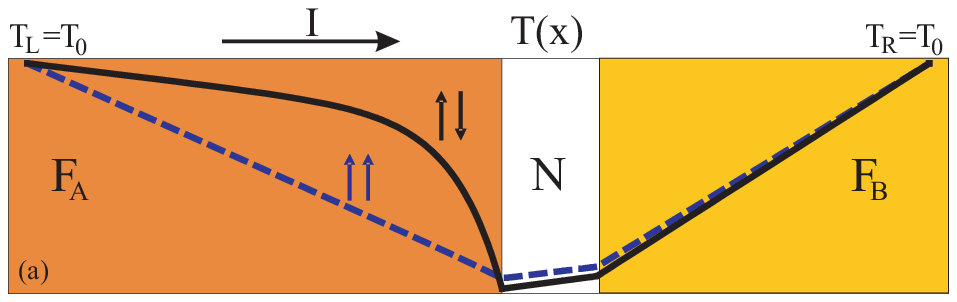}} \vspace{0.4 cm}
\resizebox{\columnwidth}{!} {\includegraphics{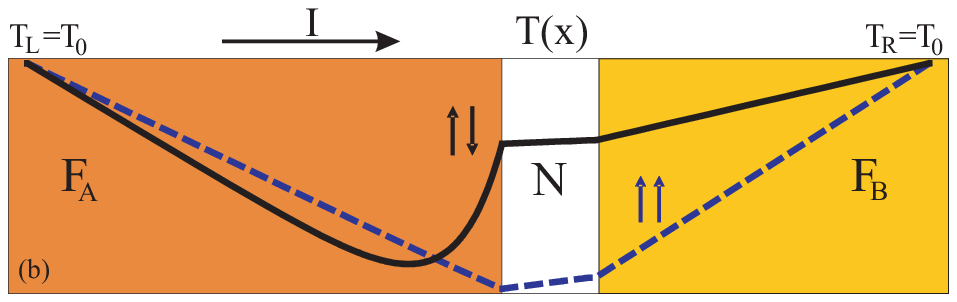}} 
\caption{Illustration of local temperature distribution in the voltage biased
$\mathrm{F_{A}|N|F_{B}}$ spin valve structure as in Fig. \ref{fig6} for both
parallel ($\uparrow\uparrow$) and anti-parallel ($\uparrow\downarrow$)
alignments of the magnetizations. The different temperature profiles
correspond to (a) $P^{\prime}\ll P$ ($P_{S}=-0.8$), (b) $P^{\prime}\gg P$
($P_{S}=-1.8$). For the parallel alignment of the magnetizations the Peltier
cooling is insensitive to spin-flips and reaches its maximum in the central
node.}%
\label{fig7}%
\end{figure}

Here we study the influence of spin-flip relaxation on the Peltier and Seebeck
effects in magnetic $\mathcal{R}_{A}|F_{A}\left(  L_{A}\right)
|N\left(  L\right)  |F_{B}\left(  L_{B}\right)  |\mathcal{R}_{B}$
nano-pillars, where $F_{A}$ and $F_{B}$ denote disordered ferromagnetic layers
, with collinear magnetization directions. 
We assume that bulk impurity scattering is dominant so that interfaces may be
disregarded. The charge and spin distribution functions in the ferromagnet,
$f_{c(s)}=\left(  f^{\uparrow}\pm f^{\downarrow}\right)  /2,$ respectively,
are then solutions of the spin diffusion equations that are continuous at the
interfaces.\cite{Kovalev:prb02} In the strongly thermalized regime, defining
$\mu_{s}=\mu_{\uparrow}-\mu_{\downarrow}$ and $\mu_{c}=(\mu_{\uparrow}%
+\mu_{\downarrow})/2$ as, respectively, the spin and charge chemical
potentials, we find (see Appendix B for details) the following thermoelectric
spin diffusion equations in a ferromagnet
\begin{align}
&  \nabla^{2}\mu_{s}=\frac{\mu_{s}}{l_{sf}^{2}},\label{spindiff}\\
&  \nabla^{2}\mu_{c}=-P\frac{\mu_{s}}{2l_{sf}^{2}},\label{chargediff}\\
&  \nabla^{2}T=\frac{(P^{\prime}-P)S}{\mathcal{L}_{0}}\frac{\mu_{s}}%
{2l_{sf}^{2}}.\label{heatdiff}%
\end{align}
Here $l_{sf}$ stands for the spin-flip diffusion length and $P$ and
$P^{\prime}$ are the spin polarizations of the bulk conductivity and its energy
derivative in the ferromagnet. These equations have to be solved with
continuity boundary conditions at the interfaces. The expressions for the
currents are similar to Eq. (\ref{pars}) after replacing
temperature and voltage differences by gradients and conductances by
conductivities. Eq. (\ref{heatdiff}) has to our knowledge not been given
elsewhere, but is required by the conservation of charge and energy currents.
According to this equation the decay of the spin accumulation in the
ferromagnet provides a source or sink of heat currents (when $P^{\prime}\neq
P$). We can understand this effect by the charge accumulation that is locally
generated by spin flips in ferromagnets, Eq. (\ref{chargediff}). Similarly,
spin-flip scattering in the presence of spin polarization of thermopower
modifies the distribution functions in a way that can be interpreted as a
source or sink of heat as expressed in Eq. (\ref{heatdiff}). The spatial
variation of $\mu_{c(s)}(x)$ and $T(x)$, in a voltage biased spin-valve is
sketched in Figs. \ref{fig6} and \ref{fig7}.
The local charge and spin chemical potentials in a spin-valve biased 
with a voltage difference (Fig. \ref{fig6}) do not depend much on the thermopower or the Peltier cooling.
More interesting are the results in Fig. \ref{fig7}, illustrating the strong dependence of 
the local temperature on magnetization configuration, the strength of the spin-flip scattering 
and the spin polarization of the thermopower.

In the following we assume identical spin polarization and spin-flip diffusion
length $l_{sf}$ in the magnetic leads and $l_{sf}^{N}\gg L$. Our results become
simple when the only asymmetries of the pillar are $\sigma_{A}\neq\sigma_{B}$
, $S_{A}\neq S_{B}$. For parallel alignments of the magnetizations the Peltier
cooling is equal to that of a normal metal structure with $R_{0}%
I_{p}(0)\approx2\mathcal{L}_{0}(G_{A}+G_{B})T\Delta T_{N}/I_{p}=2(\Pi_{A}%
-\Pi_{B})$. However, the magneto-Peltier signal in the presence of spin decay becomes
\begin{equation}
R_{0}[I_{p}(\pi)-I_{p}(0)]=-\frac{\tanh\lambda}{\lambda}\frac{4PP_S}{1+PP_S}\frac{G_{A}\Pi_{A}-G_{B}\Pi_{B}}{G_{A}+G_{B}}%
\end{equation}
where $\lambda=L_{A}/l_{sf}^{A}=L_{B}/l_{sf}^{B}$ is a measure of the
spin-flip scattering in the ferromagnets. The magneto-Peltier signal decays
with increasing $\lambda$, e.g., on using a thicker magnetic leads. The
magneto-Peltier signal vanishes when $\lambda\gg1$, and reduces to an
expression equivalent to Eq.~(\ref{MPsignal}) in the opposite limit.
Spin-flips in the normal metal spacer (with thickness comparable to or 
longer than $l_{sf}^{N}$) also reduce the magneto-Peltier signal.


For the spin-valve structure in the presence of bulk spin-diffusion in the ferromagnets, 
we find the following results for the magnetoresistance MR as well as the magnetothermopower MTP.
For the parallel configuration, as expected, no spin-flip
contribution to the total resistance and thermopower is obtained
\textit{i.e.}, $R_{P}=R_{A}+R_{B}$ and $S_{P}=(R_{A}S_{A}+R_{B}S_{B}%
)/(R_{A}+R_{B})$. However, for the anti-parallel configuration we find
\begin{align}
&  R_{AP}-R_{P}=\frac{\tanh\lambda}{\lambda}\frac{4P^{2}}{1-P^{2}}\frac
{R_{A}R_{B}}{R_{A}+R_{B}},\\
&  S_{AP}-S_{P}=-\frac{\tanh\lambda}{\lambda}\frac{2PP_S}{1+PP_S%
}\frac{R_{A}R_{B}}{(R_{A}+R_{B})^{2}}(S_{A}+S_{B})
\end{align}
The MTP is therefore proportional to the giant magneto-resistance, independent
of the spin-flip scattering strength $\lambda$. In contrast to the
MR (at $\lambda\gg1$) it is not possible to model spin-flip scattering for magneto-thermoelectric
effects by replacing $R_{A(B)}$ by the resistance of the magnetically active
region $\lambda^{-1}R_{A(B)}$.

\section{Relevance for experiments\label{exp}}

In the magnetic nanopillars considered by Fukushima \textit{et al}%
.\cite{Fukushima:jjap05,Fukushima:ieeem05} the magnetic or normal leads that
connect the central spacer to the wide external reservoirs are so long that
the bulk scattering is important: A $100\;\mathrm{%
\operatorname{nm}%
}$ long cobalt wire has a resistance $\rho_{Co}L\simeq6\;\mathrm{f}%
\operatorname{\Omega }%
\mathrm{%
\operatorname{m}%
}^{2}$ at room temperature, which is larger than the interface resistance 
$R_{Co|Cu}\approx0.25\;\mathrm{f}%
\operatorname{\Omega }%
\mathrm{%
\operatorname{m}%
}^{2}$ (see Table. I). The effective thermopower $S_{L}\approx S_{Co}\simeq-31%
\operatorname{\mu V}%
\mathrm{/}\mathrm{%
\operatorname{K}%
}$ from the bulk scattering is also larger than the interface thermopower
$S_{\mathrm{Co}|\mathrm{Cu}}\approx-6%
\operatorname{\mu V}%
\mathrm{/}\mathrm{%
\operatorname{K}%
}$. The interface contribution to the total thermopower would become more
important for high-resistance interfaces, such as tunneling barriers or point
contacts, or structures with thinner layers.

The lattice exchanges energy with the conduction electrons by inelastic
electron-phonon interactions. In principle, there is a net heat current
flowing between the electron system and the lattice/substrate. In a steady
state situation it is reasonable to assume that electron and lattice
temperatures are identical and resistance changes reflect the electron
temperature. When Peltier cooling and Joule electron heating compensate each
other the temperature change vanishes. Estimating the nonlinear electron
heating in the island by $\mathcal{L}_{0}T\Delta T_{N}=G_{tot}^{-1}%
I^{2}/2(G_{L}+G_{R})$ and using Eq. (\ref{FTN}) in terms of the lumped
conductances and thermopowers, the compensation current $I_{p}$ or 
cooling-power $R_{0}I_{p}$ for $\mathrm{F}_{A}$\textrm{%
$\vert$%
}$\mathrm{N}$\textrm{%
$\vert$%
}$\mathrm{N}_{B}$ structures such as Co$|$Cu$|$Au nanopillars can be expressed
as
\begin{equation}
R_{0}I_{p}\approx2(\gamma_{L}\Pi_{L}-\Pi_{R}).\label{peltiershift}%
\end{equation}
Such an expression holds as well for $\text{Co}|\text{Cu}|\text{Co}|\text{Au
structures}$ when the magneto-Peltier effect for the symmetric part
$\text{Co}|\text{Cu}|\text{Co}$ can be
disregarded.\cite{Fukushima:jjap05,Fukushima:ieeem05} The factor of $2$
difference with Refs.
\onlinecite{Fukushima:jjap05,Fukushima:ieeem05}
has been noted already above. We find below that including this factor leads
to a better agreement of a simple model of bulk thermopowers with experiments.
Also the too large pillar cross sections with which Gravier \textit{et
al}.\cite{Gravier:jpd06} fitted their numerical results to the experiments can
be traced back to this factor 2 in the Joule heating. Our model might not be
appropriate for Fukushima's samples that contain a highly resistant,
presumably oxide, layer over which much of the voltage drop occurs. Such a layer,
when sufficiently thick, might be better described as a bulk resistor in which
Joule heat is preferentially generated. At the compensation current $I_{p},$
finite temperature variation profiles may persist since the Joule and Peltier
sources are spatially separated. A simulation beyond our simple model might
then be required for a quantitative description.

Experimental values of $R_{0}I_{p}$ for $\text{Co}|\text{Au}$ nanopillars can
be read off the figures published by different groups, amounting to (in $%
\operatorname{mV}%
$) 19 (Ref.
\onlinecite{Albert:apl00}%
), 23.0 (Ref.
\onlinecite{Koch:prl04}%
), 22.5 (Ref.
\onlinecite{Fukushima:jjap05,Fukushima:ieeem05}%
). These numbers agree well with the following results. For a finite length of
the bulk layers, \textit{ i.e}. $L_{Co}=60\;%
\operatorname{nm}%
$ and $L_{Au}=120\;%
\operatorname{nm}%
$ and taking into account the interface scattering, when disregarding the spin
polarization of the thermopower $P_{S}$ we find $R_{0}I_{p}=19.5\;%
\operatorname{mV}%
.$  Here we also assumed $G_{R}\ll G_{L}$ caused by an oxide layer on the
non-magnetic side of the structure, $S_{\mathrm{Cu|Au}}\simeq0$,
$S_{\mathrm{Co|Cu}}\simeq-6\;%
\operatorname{\mu V}%
\mathrm{/}\mathrm{%
\operatorname{K}%
}$ and used the bulk parameters from Ref.~%
\onlinecite{Gravier:jpd06}%
. $R_{0}I_{p}\approx2(\Pi_{Co}-\Pi_{Au})=19.6\;%
\operatorname{mV}%
$ indicates that the Peltier cooling is not significantly affected by
interface scattering. A finite $P_{S}$ can enlarge or reduce the above
estimates. The spin-entropy coupling factor $\gamma_{L}\approx(1-P^{\prime
}P)/(1-P^{2})$ when the bulk and interface spin polarizations are the same.
For Co we took $P=0.44$ (Ref.
\onlinecite{Soulen:sc98}%
). Conflicting values $P_{S}=-0.18<0$ ($S^{\uparrow}=0.7S_{\downarrow}$)
(Ref.
\onlinecite{Gravier:prb06a,Gravier:prb06b}%
) and $P_{S}=+0.42>0$ ($S^{\uparrow}=-30%
\operatorname{\mu V}%
\mathrm{/}\mathrm{%
\operatorname{K}%
},S_{\downarrow}=-12%
\operatorname{\mu V}%
\mathrm{/}\mathrm{%
\operatorname{K}%
}$) (Ref.
\onlinecite{Cadeville:jpf71}%
) are found in the literature. According to Table I, $P_{S}$ of the
$\text{Co}|\text{Cu}$ interface can also have either sign. The two values for
$P_{S}$ modify the above estimate to $R_{0}I_{p}=21$ and $17%
\operatorname{mV}%
$, respectively, possibly favoring a $P_{S}<0$ when comparing with the
observed values. 

The adiabatic spin-entropy expansion term $(k_{B}T\ln2)I/e$ considered in
Ref.
\onlinecite{Katayama:jjap07}
is in our opinion an extrapolation of a concept from equilibrium thermodynamics
that does not play a role in the current induced (non-equilibrium) Peltier cooling.

We proceed by estimating the magnitude of the temperature drop that can be
realized by the Peltier effect in the magnetic heterostructure $\mathrm{Co}%
\left(  60%
\operatorname{nm}%
\right)  |\mathrm{Au}\left(  20%
\operatorname{nm}%
\right)  |\mathrm{Cr}\left(  120%
\operatorname{nm}%
\right)  $.\cite{Gravier:jpd06} At room temperature the bulk thermopowers of
both Co and Cr are relatively large and have opposite signs ($S_{\mathrm{Cr}%
}=+21.56%
\operatorname{\mu V}%
\mathrm{/}\mathrm{%
\operatorname{K}%
}$). The temperature drop $\Delta T_{\mathrm{Au}}\approx I_{p}\left(
S_{L}-S_{R}\right)  /\left(  \mathcal{L}_{0}\left(  G_{L}+G_{R}\right)
\right)  $ in the central island amounts to $4.8%
\operatorname{K}%
$ at $I_{p}=10\mathrm{%
\operatorname{mA}%
}$ for a cross-section of $70%
\operatorname{nm}%
\times200%
\operatorname{nm}%
$, at a current density of $\sim10^{8}\mathrm{%
\operatorname{A}%
/%
\operatorname{cm}%
}^{2}$, which is close to the maximum temperature drop in the temperature
profiles computed in Ref.~%
\onlinecite{Gravier:jpd06}
(we find a cooling-power $R_{0}I_{p}=30%
\operatorname{mV}%
,$ which is smaller than the observed value $41%
\operatorname{mV}%
,$ however). The temperature reduction per unit of electric current is
sensitive to the thickness of the leads. For the thick magnetic layers $\Delta
T_{\mathrm{Au}}\approx I_{p}(S_{\mathrm{Co}}-S_{\mathrm{Cr}})/\mathcal{L}%
_{0}(G_{\mathrm{Co}}+G_{\mathrm{Cr}})$. Spin polarization of the thermopower
in Co can modify the amount of the temperature reduction, up to $8\%$ for
$|P_{S}|=0.4$.

In Fukushima's experiments the leads connected to the external reservoirs are
long compared to the spin-flip diffusion length. In that regime a
magneto-Peltier effect should be small. Let us therefore consider a spin-valve
structure such as $\mathrm{Co}_{001}\left(  10%
\operatorname{nm}%
\right)  |\mathrm{Au}\left(  20%
\operatorname{nm}%
\right)  |\mathrm{Co}_{001}\left(  5%
\operatorname{nm}%
\right)  ,$ in which spin-flip scattering is less important. Due to the
different lengths of the bulk Co layers the Peltier cooling does not vanish at
this structure even for a parallel magnetic configuration; recall for example
Eq. (\ref{TNnc}) when $G_{1}\neq G_{2}$. For the parallel alignment of the
magnetizations with $P_{A}=P_{B}$ one finds $\gamma_{L(R)}^{MP}=1,$ equivalent
to a normal metal structure, whereas the spin-entropy coupling parameter
differs for the anti-parallel magnetic configuration when $P_{A}=-P_{B}$. Let
us now consider a small $G_{1}=0.01G_{2}$ \textit{e.g}. caused by an oxide
layer at the junction between the thick Co layer and the normal metal spacer,
using data in Table I for the interface scattering (at room temperature), and
adopting bulk values $P_{S}=-1.18$ and $P_{S}=0.42,$ we find respectively the
magneto-Peltier signals $R_{0}[I_{p}(\pi)-I_{p}(0)]=-1.59$ and $+6.7%
\operatorname{mV}%
$, which should be experimentally observable. Replacing the bulk parameters of
the thicker Co layer by $\rho_{Fe}=9.7\times10^{-8}%
\operatorname{\Omega }%
\mathrm{%
\operatorname{m}%
}$ and $S_{Fe}=+20%
\operatorname{\mu V}%
\mathrm{/}\mathrm{%
\operatorname{K}%
}$, the Peltier cooling is increased and the magneto-Peltier signals read 
$R_{0}[I_{p}(\pi)-I_{p}(0)]=+3.2$ and $-2%
\operatorname{mV}%
$. Finally we mention that the magneto-Peltier cooling via the bulk scattering
can be also sensitive to the degree of energy relaxation, but discussion of
the details is beyond the scope of the present paper.

Since the thermopower-to-conductance ratios $S_{i}/G_{i}$ of the intermetallic
interfaces studied up to now are smaller than the bulk values for thicker
magnetic layers, for the material combinations considered above we do not
expect an increased cooling power by reducing the thickness of the nanopillars
to the interface-dominated regime. The interface contributions are important
for (classical) point contacts or pinholes in thick tunneling barriers, since
$S_{I}$ can remain unmodified while $G_{I}$ is strongly reduced. Magnetic
tunnel junctions are interesting subjects for magneto-Peltier studies since
much higher $S/G$ ratios can be expected.

The spin Seebeck effect \cite{Uchida:nat08} recently observed in a
very long ferromagnetic metal appears to have a different origin than the
conventional mechanisms of spin and heat diffusion. The observed
thermoelectric spin signal parametrized by a spin-Seebeck coefficient
($S_{s}=-2%
\operatorname{n V}%
/%
\operatorname{K}%
$ at room temperature \cite{Uchida:nat08}) is much smaller than both the interface and bulk
thermopowers considered above. We therefore do not expect that the spin Seebeck
effect would significantly modify our findings.

\section{Summary and conclusions}

We studied the Peltier effect in nanoscale metallic multilayer structures
involving ferromagnets using a newly developed semiclassical theory of
thermoelectric transport in magnetic heterostructures including spin
relaxations and the effects of electron interactions in limiting cases. The
Peltier cooling/heating depends in general on the spin-degree of freedom as a
function of spin and energy-dependent bulk and interface scattering. We
predict a magneto-Peltier effect in spin valves, \textit{i.e}. a dependence of
Peltier cooling on the relative alignment of the two magnetization directions,
that can arise from the spin-polarization of thermopowers and is sensitive to
the spin-flip scattering as well as strength of the inelastic collisions in
the normal metal spacer. Similar behavior is found for the magneto-thermopower
which might be even easier to observe in experiments (when thermoelectric
voltage is measured rather than temperature). For ferromagnetic layers with
thickness of the order or smaller than the spin-flip diffusion length the
magneto-Peltier effect should be observable in terms of
magnetic-field-dependent resistance shifts in the $R(I)$
characteristics i.e. the cooling-power. Estimates for the Peltier cooling
based on our model and available parameters agree relatively well with
experiments as well as numerical models in which the bulk scattering dominates.

\begin{acknowledgments}
We thank J. Bass, A. Brataas, T. Heikkila, S. Maekawa, Y. V. Nazarov, S. Takahashi,
J. Xiao for helpful discussions. This work is supported by
``NanoNed'', a nanotechnology programme of the Dutch Ministry of
Economic Affairs. It is also part of the research program for the
``Stichting voor Fundamenteel Onderzoek der Materie'' (FOM) and the use
of supercomputer facilities was sponsored by the ``Stichting Nationale
Computer Faciliteiten'' (NCF), both financially supported by the
``Nederlandse Organisatie voor Wetenschappelijk Onderzoek'' (NWO).
\end{acknowledgments}

\appendix

\section{Phonons}

The Peltier effect in the presence of phonon heat conduction and
electron-phonon interactions can be modeled in linear response as follows: The
net heat current flowing between the electron and phonon subsystems of the
island for small temperature differences $T_{e}^{N}-T_{p}%
^{N}\ll T_{p}^{N}$ may be parametrized by the simple linear equation $\dot
{Q}_{e-p}=-\kappa_{e-p}(T_{e}^{N}-T_{p}^{N})$. \cite{Groeneveld:prb95} For a
phonon temperature drop of $\Delta T_{p}$ across an interface, $\dot{Q}%
_{p}=-\kappa_{p}\Delta T_{p}$ with $\kappa_{p}$ the phonon thermal conductance
of the junction. \cite{Gundrum:prb05} The energy conservation laws then read:
$\dot{Q}_{e1}+\dot{Q}_{e2}+\dot{Q}_{e-p}=0$ and $\dot{Q}_{p1}+\dot{Q}%
_{p2}-\dot{Q}_{e-p}=0$ for the electron and phonon subsystems, respectively.
The electron temperature in the node, Eq. (\ref{TN1}), is then modified as follows
\begin{equation}
\Delta T_{e}^{N}=\frac{(\Pi_{1}-\Pi_{2})I}{\kappa_{e1}+\kappa_{e2}+\gamma
_{p}(\kappa_{p1}+\kappa_{p2})}%
\end{equation}
and $\Delta T_{p}^{N}=\gamma_{p}\Delta T_{e}^{N}$ where $\gamma_{p}%
=\kappa_{e-p}/(\kappa_{p1}+\kappa_{p2}+\kappa_{e-p})$. In the limit
$\kappa_{p1(2)}\ll\kappa_{e-p}$ the Peltier cooling is reduced by the sum of
the total thermal conductances $\kappa_{1(2)}=\kappa_{e1(2)}+\kappa_{p1(2)}$.
The figure of merit $S\Delta T_{e}^{N}/\Delta V$ is then further decreased by
taking into account the contribution of the phonon heat conduction ($\dot
{Q}_{p}\neq0$).

\section{Thermoelectric spin diffusion equations}

In a diffusive magnetic metal in the steady state the Boltzmann transport
equation in the relaxation time approximation leads to the following spectral
spin diffusion equations for the local variation of the spin distribution
functions $f^{\uparrow(\downarrow)}(\epsilon)$ for each spin $\alpha$ as
\begin{equation}
\nabla^{2}f^{\alpha}(\epsilon)=\frac{f^{\alpha}(\epsilon)-f^{-\alpha}%
(\epsilon)}{(l_{sf}^{\alpha})^{2}},
\end{equation}
where $l_{sf}^{\alpha}=\sqrt{D^{\alpha}\tau_{sf}^{\alpha}}$ are the
spin-dependent diffusion lengths. Under the detailed balance condition
$\mathcal{N}^{\uparrow}/\tau_{sf}^{\uparrow}=\mathcal{N}^{\downarrow}%
/\tau_{sf}^{\downarrow}$ the spectral spin diffusion equations can be
rewritten as
\begin{align}
&  \nabla^{2}f_{s}(\epsilon)=\frac{f_{s}(\epsilon)}{l_{sf}^{2}},\\
&  \nabla^{2}f_{c}(\epsilon)=-P\frac{f_{s}(\epsilon)}{2l_{sf}^{2}}.
\end{align}
where $(l_{sf})^{-2}=(l_{sf}^{\uparrow})^{-2}+(l_{sf}^{\downarrow})^{-2}$ and
the charge and spin distribution functions, $f_{c(s)}=\left(  f^{\uparrow}\pm
f^{\downarrow}\right)  /2,$ have been introduced. In the strongly thermalized
regime the spin diffusion equations can be expressed in terms of the spin
chemical potentials $\mu_{\uparrow(\downarrow)}=(\mu_{c}\pm\mu_{s}/2)$ and the
electron temperature $T$. After inserting the linear expansions
\begin{align}
&  f_{s}(\epsilon)\approx\left(  -\frac{\partial f_{0}}{\partial\epsilon
}\right)  \mu_{s},\\
&  \nabla^{2}f_{c(s)}(\epsilon)\approx\left(  -\frac{\partial f_{0}}%
{\partial\epsilon}\right)  \left(  \nabla^{2}\mu_{c(s)}+\frac{\epsilon-\mu}%
{T}\nabla^{2}T\right)  ,
\end{align}
into the above diffusion equations we can integrate over energies by using the
Sommerfeld approximation (see Eq. (\ref{S-int})). We assume $S^{2}%
\ll\mathcal{L}_{0}$ and disregard an energy dependence of the spin diffusion
length $l_{sf}$ (which is allowed when $2P S_{sf}\ll (P^{\prime}-P)S$ in which 
$S_{sf}=-eL_0 T \partial_{\epsilon}ln l_{sf}(\epsilon)|_{\epsilon_F}$), 
but keep the energy dependence of the spin polarization $P$, recall Eq. (\ref{P-energy}).
One then arrives at the thermoelectric spin diffusion equations expressed in
Eqs. (\ref{spindiff}-\ref{heatdiff}). Among the spin diffusion equations Eqs.
(\ref{spindiff}) and (\ref{chargediff}) are already well
known.\cite{Valet:prb93} Eq. (\ref{heatdiff}) represents a spin-heat coupling
for the electron spin diffusion in the presence of spin polarization of thermopower.



\end{document}